%% file: dcm15.tex
\documentclass[submission,copyright,creativecommons]{eptcs}

\usepackage[latin1]{inputenc}
\usepackage{amsmath}
\usepackage{amssymb}
\usepackage{amsthm}
\usepackage{enumerate}
\usepackage{xspace}
\usepackage{amsfonts}
\usepackage{mathrsfs}
\usepackage[all,2cell]{xy}
\CompileMatrices 
\UseAllTwocells
\SilentMatrices
\usepackage{graphicx}
\usepackage{cite}
\usepackage{float}
\usepackage{fancybox}

\input macro.tex

\title{Compositional model checking of concurrent systems, with Petri nets}
\author{Pawe{\l} Soboci{\'n}ski
\institute{ECS, University of Southampton, UK}}

\def\authorrunning{P.\ Soboci\'{n}ski}
\def\titlerunning{Compositional model checking of concurrent systems, with Petri nets}

\newcommand{\cgr}[2][scale=0.45]{\raisebox{0.1em}{\begingroup
\setbox0=\hbox{\includegraphics[#1]{#2}}%
\parbox{\wd0}{\box0}\endgroup}}

\begin{document}
\maketitle

\begin{abstract}
Compositionality and process equivalence are both standard concepts of process algebra. Compositionality means that the behaviour of a compound system relies only on the behaviour of its components, i.e. there is no emergent behaviour. Process equivalence means that the explicit statespace of a system takes a back seat to its interaction patterns: the information that an environment can obtain though interaction. 

Petri nets are a classical, widely used and understood, model of concurrency. Nevertheless, they have often been described as a non-compositional model, and tools tend to deal with monolithic, globally-specified models. 

This tutorial paper concentrates on Petri Nets with Boundaries (PNB): a compositional, graphical algebra of 1-safe nets, and its applications to reachability checking within the tool \texttt{Penrose}.
The algorithms feature the use of compositionality and process equivalence, a powerful combination that can be harnessed to improve the performance of checking reachability and coverability in several common examples where Petri nets model realistic concurrent systems.
\end{abstract}

\section{Introduction}
This short paper is a tutorial on the algebra of Petri nets with boundaries (PNB) and its use as a compositional formalism for the modelling of concurrent and distributed systems, and their compositional verification via the tool \texttt{Penrose}. While compositional approaches to Petri nets have a long history---starting with Mazurkiewicz~\cite{Mazurkiewicz1988}---compositional reasoning has not had a significant impact in modelling approaches and tools. Indeed,  \texttt{Penrose} is the first automated reachability checker to harness compositionality; more established tools deal with the global statespace. 

This article is deliberately non-technical and gives a high-level overview of the work on PNB. The reader hungry for more detail will find it, together with pointers to the significant corpus of related work, in the cited research papers. A comprehensive treatment can be found in Stephens' recent PhD thesis~\cite{Stephens2015}.

The paper is divided into three sections: Section~\ref{sec:pnb} is an overview of the algebra of PNB. Section~\ref{sec:penrose} is devoted to an explanation of how the algebra underpins a compositional algorithm for checking reachability via the \texttt{Penrose} tool. In Section~\ref{sec:future} we outline the main lines of anticipated future work. The reader is assumed to already be familiar with the basic concepts of Petri nets and their \emph{step} semantics where independent transitions can be fired in parallel.


\section{Compositional modelling: Petri nets with boundaries}\label{sec:pnb}


Petri nets with boundaries (PNB) are a compositional algebra of Petri nets. The algebra was introduced for 1-safe nets---where the step semantics ensures that there is at most one token on each place by disabling the firing of those transitions that would violate this invariant---in the CONCUR 2010 paper by the author~\cite{Sobocinski2010}. Subsequently, it was extended by Bruni, Melgratti and Montanari for ordinary, infinite-state place/transitions nets in their CONCUR 2011 paper~\cite{Bruni2011}. The theory was then consolidated and presented together in a uniform manner in the jointly authored archival version~\cite{Bruni2013}. 

To understand the basic idea of compositionality in the context of Petri nets with boundaries, it is instructive to think of a Petri net as a system of \emph{actors} (the places) that may participate in \emph{distributed synchronisations} (the transitions). 
Given a particular synchronisation, it is then possible to take a local view from the point of a subset of the involved actors. Indeed, the local information relevant to a single actor is: 
\begin{enumerate}
\item whether it is able to participate in the synchronisation, i.e.\ does it satisfy the requirements of the Petri net transition?
\item if so, what is the effect of the synchronisation on its internal state: i.e.\ what is the effect of firing the transition on its number of tokens?
\end{enumerate}
Let us illustrate these points on the most trivial relevant example: a 1-safe Petri net with a single transition that takes a token from place $A$ and produces one at place $B$. 
\begin{equation}\label{eq:firstExample}
\cgr[height=1.5cm]{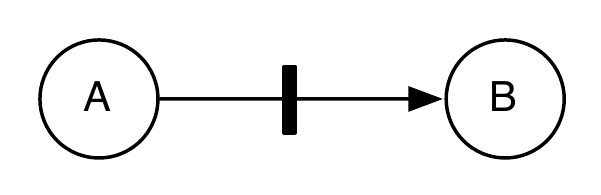}
\end{equation}
Here we can think of $A$ and $B$ as simple computational entities that can be in one of two states: $0$ or $1$ depending on whether a token is present. In order for the synchronisation---the firing of the transition in~\eqref{eq:firstExample}---to occur, $A$ must contain a token and $B$ must be empty. The effects of the synchronisation are to remove the token from $A$ and place a token on $B$. 

This information is all we need to consider two local views of the synchronisation: ``$A$'s'' view and ``$B$'s'' view. First, we need to separate the transition into two components.
\begin{equation}\label{eq:decomposition}
\cgr[height=2cm]{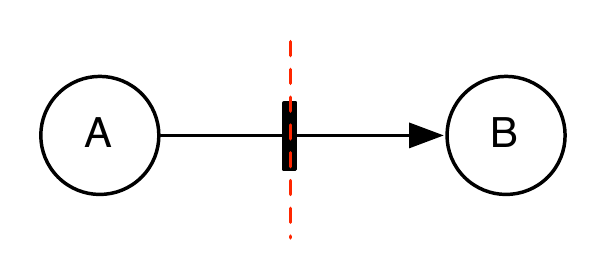}
\end{equation}
The red dashed line in the diagram above represents a \emph{boundary} between the two components. As evident in the diagram, the synchronisation passes through the boundary. The basic insight is now that the global synchronisation can be modelled by performing it piecewise in the two components, while synchronising the two actions on the boundary.
The local effect of each component can be conveniently modelled with simple labelled transition systems (LTS), for $A$ and $B$, respectively, reading from left to right.
\begin{equation}\label{eq:transitionSystems}
\cgr[height=3.5cm]{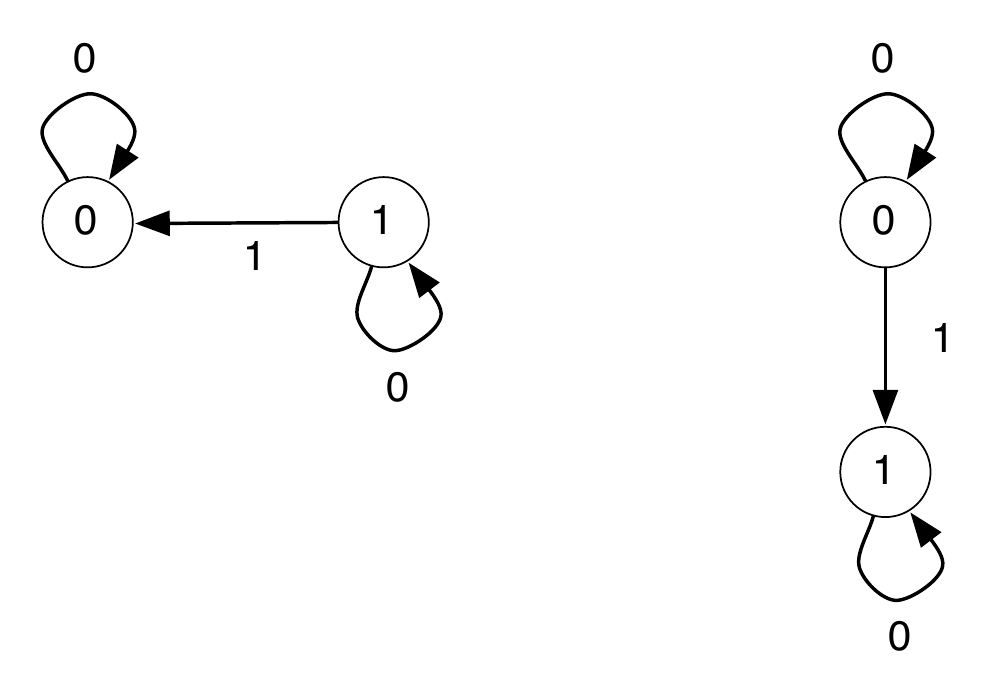}
\end{equation}
The circles in~\eqref{eq:transitionSystems} are now not Petri net places, but individual states of $A$ and $B$: state labelled $0$ means the state where there is no token, $1$ means a single token. The labels on the transitions, although also labelled with $0$s and $1$s, refer to something quite different: the capability of changing the internal state of the component, while possibly engaging in synchronisation. Whether or not a state transition engages in a synchronisation is represented by the label of the transition, and can be thought of as \emph{what can be observed on the boundary}. 

The label $0$ means that no synchronisation is engaged.  As witnessed by the loop transitions on each state, each component, in each state, has the possibility of not changing state by not engaging in any transitions. Indeed, the transition systems that arise from PNB are always \emph{reflexive} in this sense: every LTS state has a self-looping $0$ labelled transition. The label $1$ means that the synchronisation has been engaged in exactly once. 
For example, in the left transition system of \eqref{eq:transitionSystems} that corresponds to place $A$, the $1$ labelled transition records the fact that when the transition fires the place loses its token: that is, there is a transition from state $1$ to state $0$.

In the 1-safe case, there can be no auto-concurrency on transitions and thus the LTS transition alphabet is limited to the set $\{0,1\}$. In the infinite state case, however, it is natural to consider the labels coming from the set of the natural numbers, since individual transitions can, in principle be fired simultaneously: see~\cite{Bruni2013} for the details.

Now to obtain the global step semantics of the original Petri net~\eqref{eq:firstExample} one needs to synchronise the two transition systems. The synchronisation is a natural variant of the cartesian product of the two transition systems: the resulting states are simply ordered pairs of states of the original LTSs,
and a transition $(a,b)\to (a',b')$ exists
whenever we can find $a\xrightarrow{\alpha}a'$ and $b\xrightarrow{\alpha}b'$ in the original transition systems.
The fact that $\alpha$ is the same label in both component transitions means that we are \emph{synchronising the two on the common boundary}. 
\[
\cgr[height=4.5cm]{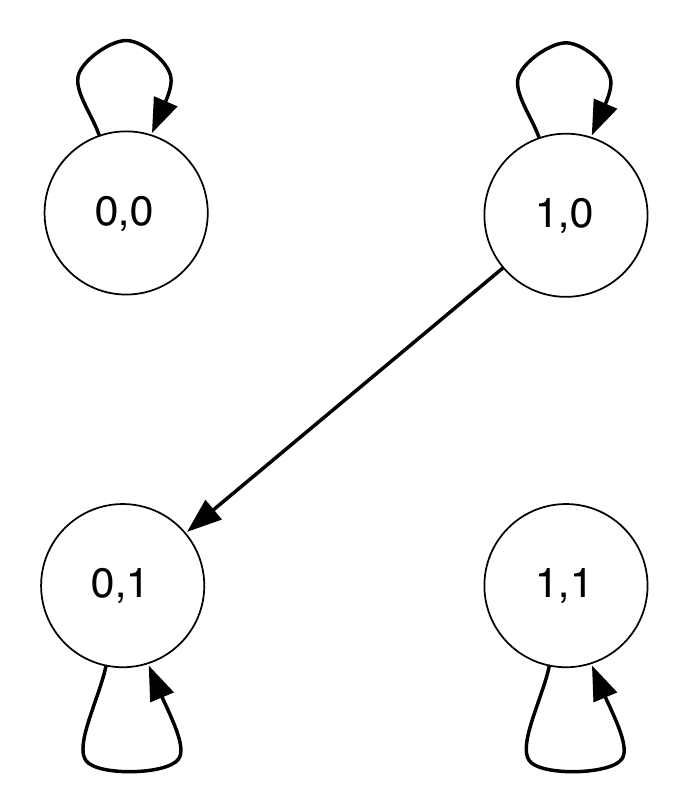}
\]
The above transition system represents the result of the synchronisation: we see that, as expected, the only non-trivial behaviour of the net~\eqref{eq:firstExample} is when place $A$ contains one token and place $B$ zero tokens: in that case, the transition can be fired, resulting in a transition to the state where the token placement has been inverted.

\medskip
The basic ideas outlined above can be distilled into two observations:
\begin{enumerate}
\item net transitions, seen as a distributed synchronisations, can be separated into individual components;
\item the seemingly fundamental classification of \emph{inputs} and \emph{outputs} of a particular net transition is actually \emph{local} --- whether a synchronisation acts as an input (token consumer) or an output (token producer) is something that is determined by the individual component's transition system.  
\end{enumerate}

These observations leads to the notion of \emph{algebra} of components, which consists of two operations that allow us to connect them into composite, interacting networks. They also suggest an alternative graphical notation that makes the input/output assignment of transition endpoints local; it is the graphical syntax adopted in the literature on PNBs and we will use it henceforward.
The net~\eqref{eq:firstExample}, using the alternative graphical syntax, is pictured below.
\begin{equation}\label{eq:newSyntax}
\cgr[height=1.5cm]{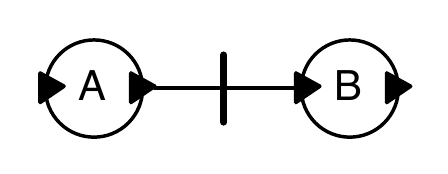}
\end{equation}
Each place now has two \emph{ports}: an \emph{input} port, drawn as a black triangle pointing into it, and an \emph{output} port, represented by the black triangle pointing out. A transition is determined by a subset of the set of all ports of a particular net: in~\eqref{eq:newSyntax} the transition is the set containing the output port of $A$ and the input port of $B$. The vertical line through the transition is a notational convenience that helps to distinguish between different transitions.

Roughly speaking, rather than directing transitions, we direct the places. It is not difficult that the two graphical representations have equal expressive power and we can readily translate between them. The alternative graphical syntax comes into its own when describing net components, because it frees one from the red-herring distraction of directed transitions and the implicit type information it brings: for example, it is no problem to, for instance, connect an ``input to an input'': the result is that two partial views of a synchronisation are combined, and the two local input actions are synchronised.

The intuitive idea of separating the simple synchronisation~\eqref{eq:decomposition} into its constituents has a formal status as a PNB expression~\eqref{eq:expression}. Its two components are illustrated below.
\begin{equation}\label{eq:components}
\cgr[height=2cm]{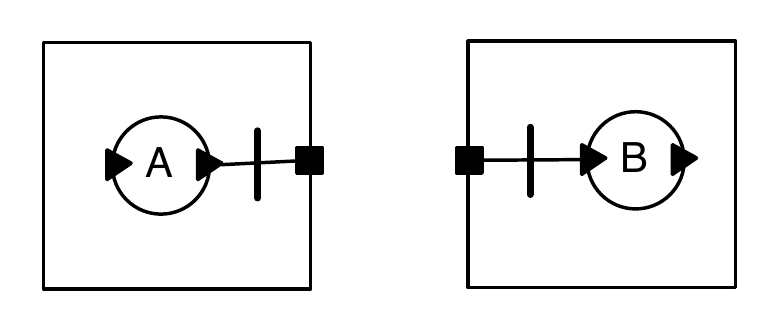}
\end{equation}
Here the common boundary is made explicit: it is represented by the \emph{boundary ports}: one on the right hand side of the box that encloses the place $A$ and the second the left hand side of the box around place $B$. A boundary port plays a similar role to the input and output ports on places in the sense that transition can connect to them: here the original transition connecting $A$ and $B$ has been split into a transition that connects the output port of $A$ with the right boundary port in the first component, and a transition that connects the input port of $B$ with the left boundary port in the second. Note that, although ports can appear either on the left or on the right, they ought not be confused with inputs and outputs since they act merely as synchronisation points; as previously mentioned, inputs and outputs are notions local to places. Indeed, one could just as well use the components below in a decomposition of~\eqref{eq:newSyntax}; intuitively, now, the flow is from right to left.
\[
\cgr[height=2cm]{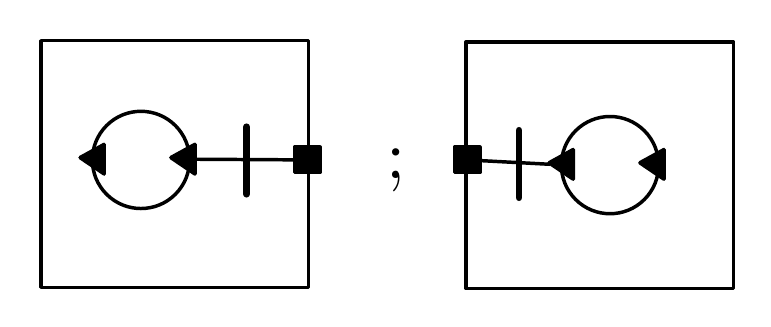}
\]
 In terms of a physical intuition, the boundary ports play a role akin to \emph{terminals} in engineering terminology; the idea is that a system interacts with its environment via its terminals.

There are two operations with which one can compose PNB components. The first is \emph{composition along a common boundary}. Given a PNB $M$ of type $(m,k)$, where $m,k\in\N$ are the numbers of, respectively, the left hand side ports and the right hand side ports, and a PNB $N$ of type $(k,n)$, we obtain a PNB $M\mathrel{;}N$ of type $(m,n)$.
The places of $M\mathrel{;} N$ are the disjoint union of the places of $M$ and $N$; its transitions are the union of the transitions of $M$ and $N$ which do not connect to the common boundary, together with \emph{synchronisations} of those transitions that do connect to boundary ports. For example, composing the two components in~\eqref{eq:components} of 
type $(0,1)$ and $(1,0)$, respectively, yields the PNB of type $(0,0)$ illustrated below.
\begin{equation}\label{eq:expression}
\cgr[height=2.5cm]{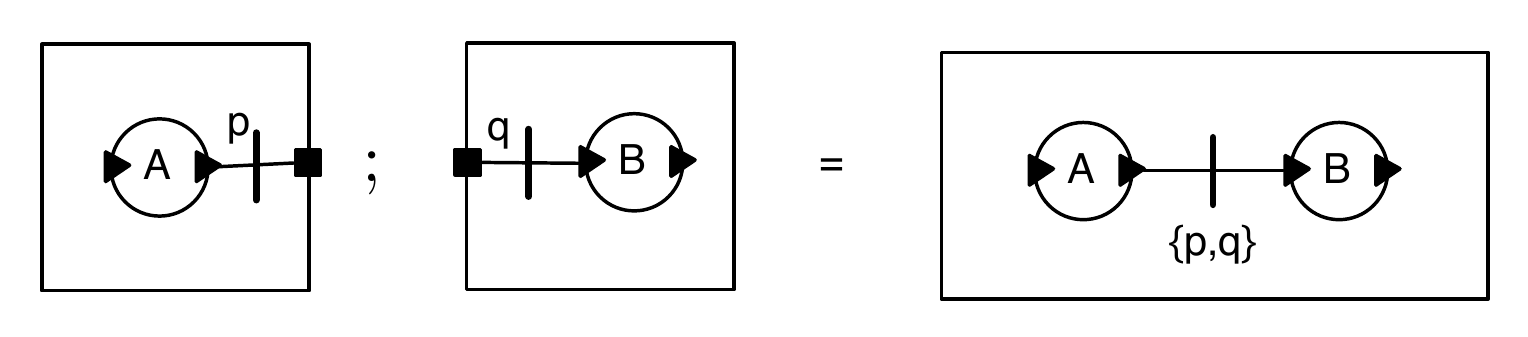}
\end{equation}
The transition $\{p,q\}$ in the result is the synchronisation of $p$ and $q$ in the constituent components.
Rather than dwell on the formal details of the definition of composition, which can be found in~\cite{Bruni2013}, we illustrate it with a number of examples which give the intuition. 



\[
\cgr[height=6cm]{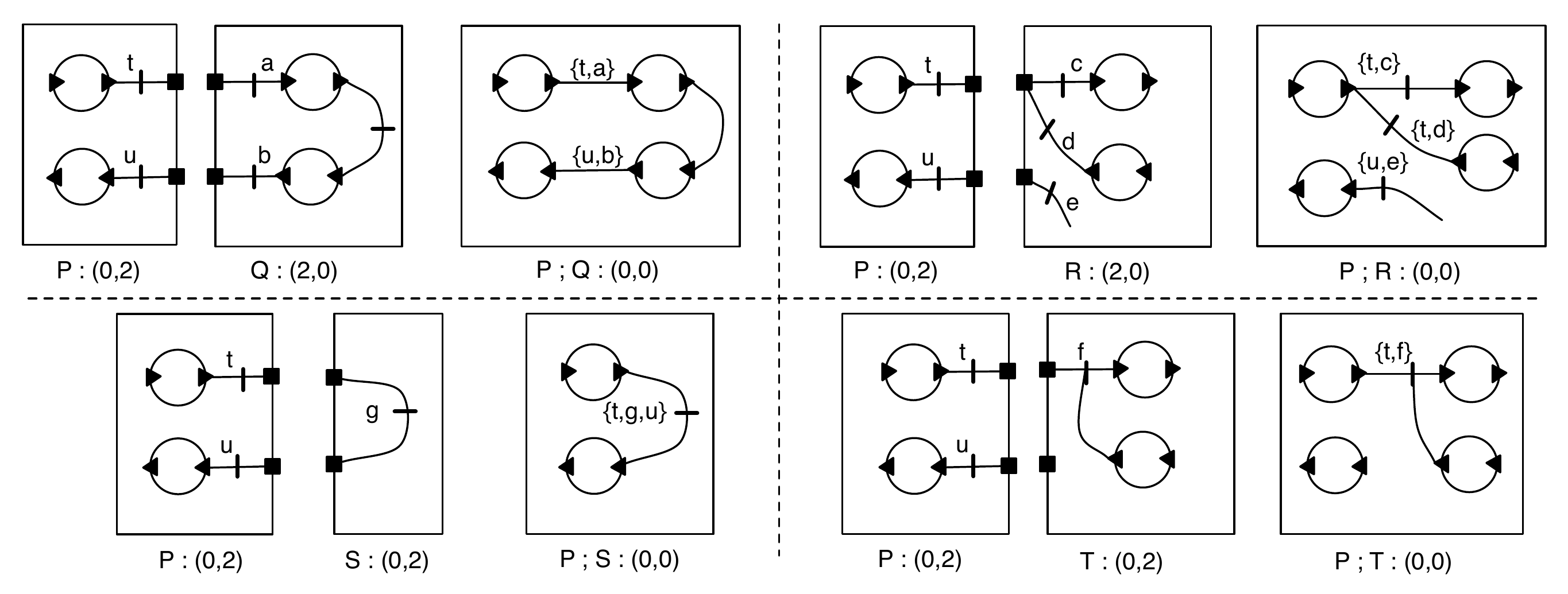}
\]
Starting with the top left example, we see that a composite net is obtained by synchronising transitions on two separate ports. Note that the order of the ports is significant. The top right example contains two interesting situations: first,  the two transitions $c$ and $d$ of component $R$ both connect to the upper left hand side port. In the composition, transition $t$ can synchronise with either and both the possibilities, $\{t,c\}$ and $\{t,d\}$ are transitions in the resulting net. The synchronisation on the second port is also of interest: here $e$ is a transition that connects only to the port. The result of synchronising $u$ with $e$ is that $u$ is completed; and becomes a transition which does not have any effect on the places coming from the right component. While seemingly spurious, transitions that connect only to boundary ports are actually very useful in examples, as we will see in Section~\ref{sec:penrose}.
In the lower left example, we see that the composition has the effect of synchronising all three of the transitions $t$, $u$ and $g$, even though $t$ and $u$ were not originally synchronised. This is because the transition $g$ is connected to both of the boundary ports: thus, to form a part of a transition in the result it needs to synchronise with corresponding transitions that connect to the shared boundary ports in the left hand side component. Finally, the lower right example ought to be contrasted with the upper right: here there is a single transition $f$ that connects to the upper left boundary port and the two places. The result in the composition is a single synchronisation $\{t,f\}$.
On the other hand, $u$ now has no partner transition in $T$ with which it can synchronise, therefore there is no transition in the composition that has $u$ as a component.

The second operation of the algebra of PNB is a non-interacting parallel composition. Given 
nets $M$ of type $(m_1,n_1)$ and $N$ of type $(m_2,n_2)$, their parallel composition $M\oplus N$ has
type $(m_1+m_2,n_1+n_2)$. This operation is much easier to describe: the resulting net is simply the
disjoint union of the two nets, where the left and right ports of $M$ come before (ordering from top to bottom)
the respective ports of $N$. Graphically, this can be understood simply as ``stacking'' $M$ on top of $N$. This operation is very useful in constructing cyclic networks (e.g.\ a token ring) as well as nets with repeated structure, such as trees, see~\cite{Rathke2013,Stephens2015} for examples. The simplicity of the examples in this tutorial means that we will not need to use it further.

The passage from a net component to its transition system, as previewed in the move from~\eqref{eq:decomposition} to~\eqref{eq:transitionSystems} gives us a notion of semantics of PNB. The semantics is a two-labelled transition system (2LTS); each transition has two labels. For instance, let us focus the semantics of the PNB pictured below.
\begin{equation}\label{eq:semanticsNet}
\cgr[height=1.5cm]{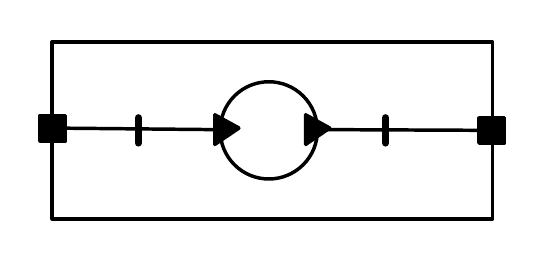}
\end{equation}
Its transition system, pictured below, has as its states the possible markings of the net; in the particular case of~\eqref{eq:semanticsNet} there are just two possible markings, so two states.
\begin{equation}\label{eq:semanticsNetLTS}
\cgr[height=4cm]{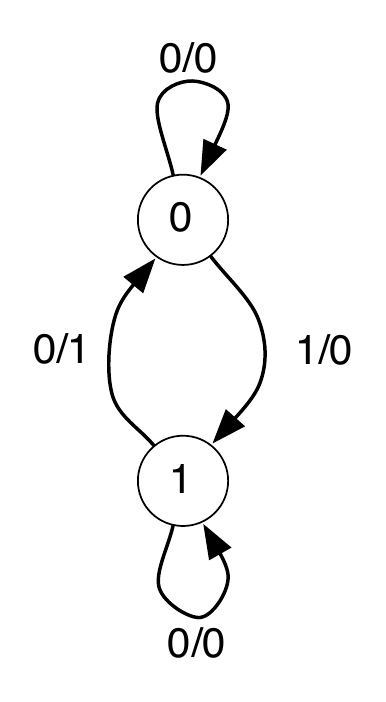}
\end{equation}
As mentioned previously, each transition of~\eqref{eq:semanticsNetLTS} has two labels: the first to the left of the `/' separator and the second to the right. The left label refers to the interactions on the left boundary, and similarly, the right to the interactions on the right. In particular, at the empty marking (state $0$), the net can synchronise on the left, producing a token, that is transitioning to state $1$. From that state, it can synchronise on the right to come back to the empty marking. In any state there is the option of not synchronising and remaining in the same state. In general, the left and right labels are \emph{words}
over the alphabet $\{0,1\}$, their length determined by the size of the boundary. Specifically, a net $M$ of type $(m,n)$ will have all of its transitions having labels of the form $\alpha/\beta$ where $\alpha\in\{0,1\}^m$ and
$\beta\in\{0,1\}^n$. The general procedure of obtaining a 2LTS from a PNB can be seen as a straightforward generalisation of the familiar notion of step firing semantics.

There is also a straightforward way of defining the operations of $\mathrel{;}$ and $\oplus$ \emph{directly} on 2LTS, and both are variants of the product of transition systems. For example, the rule
\begin{equation}\label{eq:compositionRule}
\frac{a \xrightarrow{\alpha/\beta} a' \quad b\xrightarrow{\beta/\gamma}b'}
{(a,b) \xrightarrow{\alpha/\gamma} (a',b')}
\end{equation}
defines the composition `$\mathrel{;}$' and witnesses our intuition of this operation as synchronisation along a shared boundary.

Compositionality of the formalism can now be expressed precisely: 
given nets $M$ of type $(m,k)$ and $N$ of type $(k,n)$, to obtain the semantics of $M\mathrel{;}N$
we can either
\begin{itemize}
\item translate $M$ and $N$ individually into transition systems and compose them using~\eqref{eq:compositionRule};
\item compose $M$ and $N$ as PNBs and obtain a transition system for the composite net, using the step firing semantics. 
\end{itemize}
Compositionality means that it does not matter which choice is made: the two transition systems are guaranteed to be isomorphic.

We conclude this section with a brief observation on the category theoretic account of compositionality, in the sense described above. Petri nets with boundaries form the arrows of a prop: a strict symmetric monoidal category with object the natural numbers, where the monoidal product on objects is addition. Next, two-labelled transition systems are also the arrows of a prop. Then compositionality, stated succinctly, is that the translation from nets to transition system is a prop homomorphism.


\section{Compositional reachability checking}
\label{sec:penrose}


Our running example for this section is a simple, modular specification of a counter Petri net\footnote{This example was prompted by a question by Moshe Vardi during the DCM workshop in Cali, Colombia, in October 2015.} using the algebra of PNB.
An $n$-bit counter consists of $n$ basic components, each modelling the behaviour of an individual bit. The component that models the behaviour of each bit is illustrated below.
\[
\cgr[height=2.5cm]{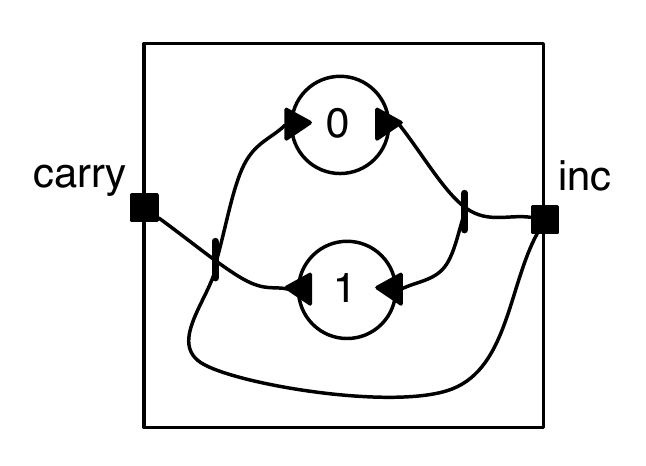}
\]
Assuming a token on place zero, a synchronisation on the right boundary port (marked \textsf{inc}) flips the bit to $1$, that is, the token is removed from place $0$ and placed at $1$. Another synchronisation on port \textsf{inc}, enabled when there is a token at place $1$, takes us to the starting state, but the synchronisation now also involves the left hand side \textsf{carry} port, which in turn affects the state of higher order bits. For example, composing three of these components in series gives us a $3$-bit counter, as illustrated below. In the graphical representation of the result of the composition, colours and various line styles have been used in order to help distinguish between the transitions.
\begin{equation}\label{eq:3bitComposition}
\cgr[height=3cm]{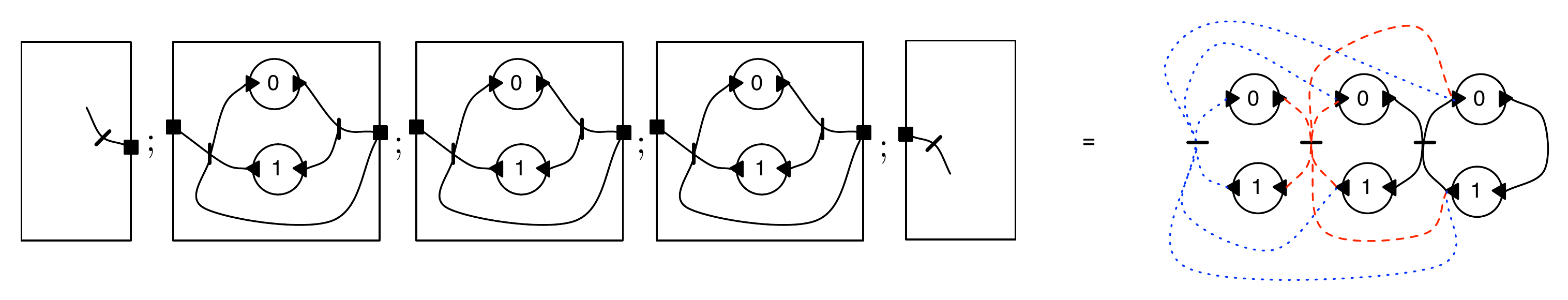}
\end{equation}
One natural question to ask about the $n$-bit counter net in general is whether the state in which all of the bits are $1$ is reachable from the state in which they are all in state $0$. 
Traditionally, algorithms for deciding reachability consider the global state space, but in our counter example, the shortest firing sequence from the initial state to the desired state has length $2^n-1$. This is not surprising; indeed, reachability checking for safe nets is PSPACE-complete~\cite{Cheng1993}.

Here the compositional approach can be useful. First, note that the reachability question itself is inherently local in the sense that the reachability problem leads us to consider each component's transition system as a nondeterministic finite automaton (NFA): the starting state being the initial marking and the final state(s) the desired marking(s). This idea is illustrated below: starting with the transition system for a counter net component, illustrated on the left, we can associate the reachability specification and consider it as an automaton. Indeed, the initial state is when the bit is in state $0$, and the final state $1$ is the desired marking.
\begin{equation}\label{eq:componentAut}
\cgr[height=3.5cm]{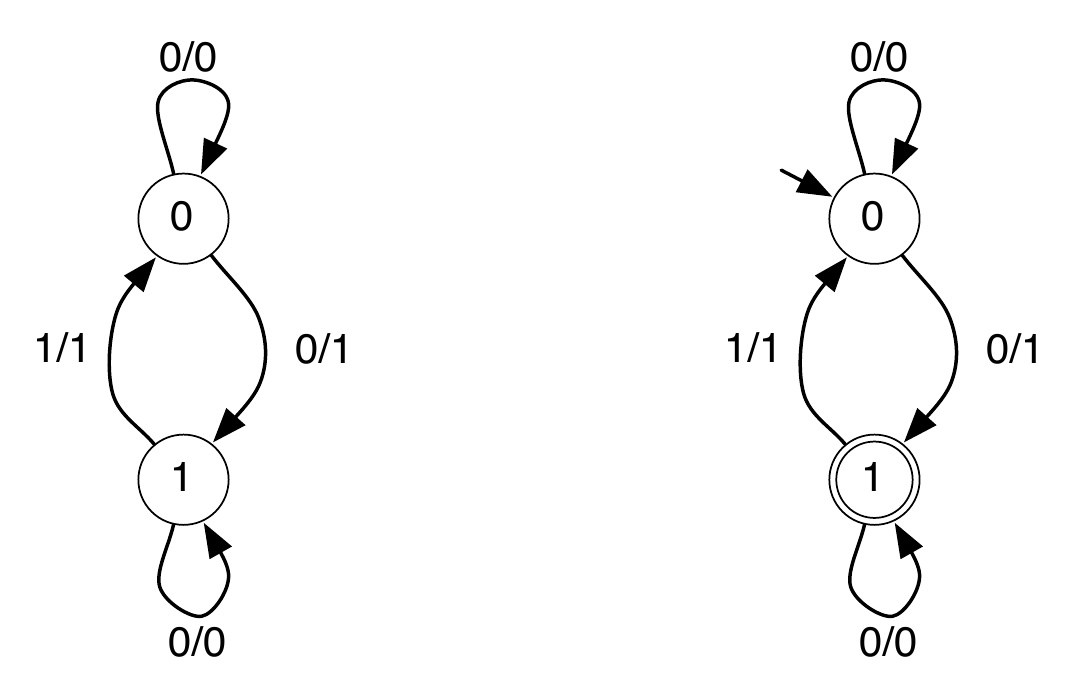}
\end{equation}
Clearly, the reachability question of the $n$-bit counter net reduces to language emptiness of the composed, global automaton that is the semantics of the composed net. This automaton has $2^n$ states, so by itself, it is not very useful. Here \emph{compositionality combined with language equivalence} does the trick. To illustrate this point, let us consider the following composition, the rightmost part of~\eqref{eq:3bitComposition}.
\begin{equation}\label{eq:rightTail}
\cgr[height=2.5cm]{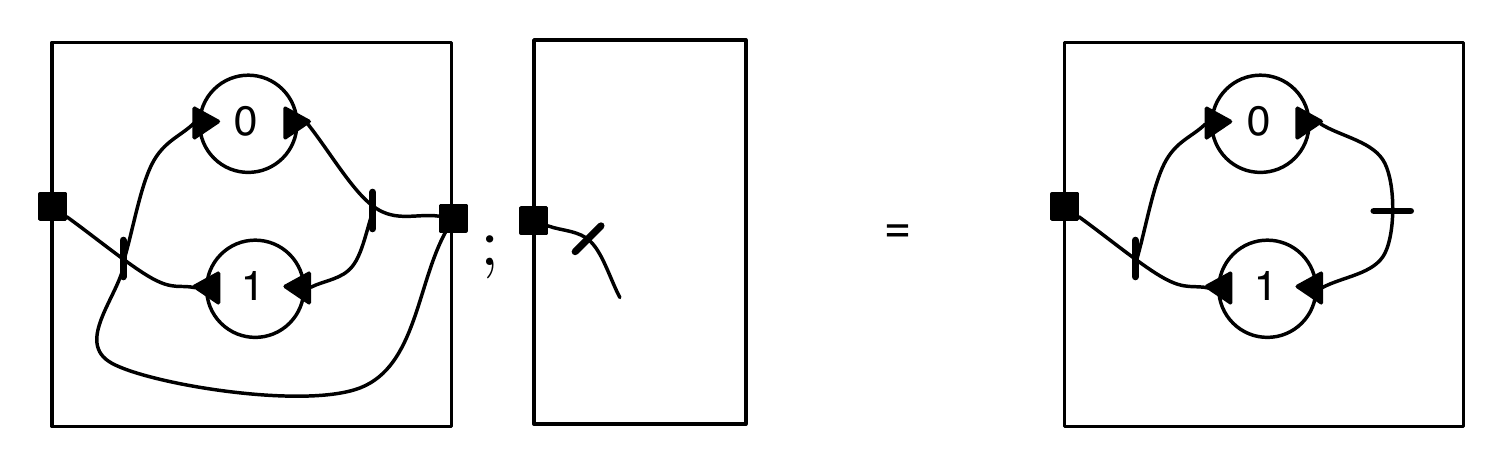}
\end{equation}
Using compositionality, the automaton for~\eqref{eq:3bitComposition} can be obtained by composing~\eqref{eq:componentAut} with the automaton
\begin{equation}\label{eq:rightEndAut}
\cgr[height=2.5cm]{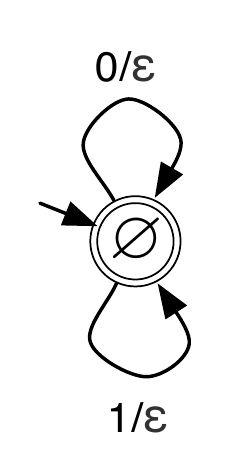}
\end{equation}
which is the semantics of the rightmost component in the composition~\eqref{eq:3bitComposition}. The result is the following:
\begin{equation}\label{eq:rightEndComposed}
\cgr[height=3.5cm]{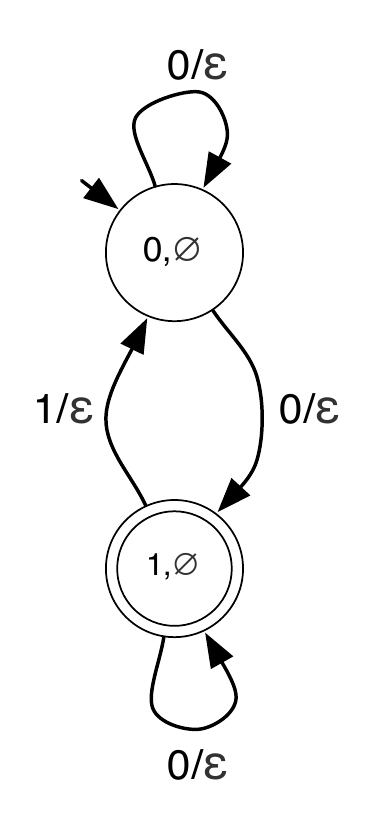}
\end{equation}
At this point, it is instructive to take a step back and discuss further the role of the \emph{zero transitions}. By zero transitions we mean those where the component labels are strings made up solely of $0$s. We have seen, starting with~\eqref{eq:transitionSystems} that the transition systems that underlie PNB components are reflexive: at each state there is a looping zero transition. But not all zero transitions are self loops; for example, in the transition system of~\eqref{eq:rightEndComposed}, there is $0/\epsilon$ transition from state $(0,\varnothing)$ to state $(1,\varnothing)$; it corresponds to the firing of an \emph{internal} transition---that is, one not connected to a boundary port---of the component that is the result of the composition in~\eqref{eq:rightTail}.

The fact that zero transitions are internal means that they can be safely ignored when reasoning about the behaviour of composed nets. More precisely, \emph{weak language equivalence} is a congruence wrt to the operations ($\oplus$ and $;$) of PNBs~\cite[Proposition 8]{Rathke2014a}. Weak language equivalence is closely related to the classical notion of $\epsilon$-closure in NFAs; zero transitions are treated akin to $\epsilon$-transitions. This means that we can replace~\eqref{eq:rightEndComposed} with any (smaller) weak language equivalent automaton and use it in subsequent compositions, without sacrificing correctness wrt reachability checking. Indeed, the automaton~\eqref{eq:rightEndComposed} can be replaced with the weakly equivalent~\eqref{eq:rightEndAut}.

At this point, we have a proof of reachability since adding additional counter components will result in~\eqref{eq:rightEndAut}. This procedure has been automated, resulting in the tool \texttt{Penrose}, as detailed in~\cite{Rathke2014a, Stephens2015}. The input to the tool is a PNB expression, such as~\eqref{eq:3bitComposition}, viewed as a syntax tree where the leaves are labelled by individual PNB, and the internal nodes by the PNB operations $\oplus$ and $\mathrel{;}$. The computation done by the tool can then be summarised as consisting of the following steps:
\begin{enumerate}[(i)]
\item Translate the leaves of the expression to automata, then evaluate, bottom up, as follows.
\item In order to compose automata $A$ and $B$, first minimise them 
wrt weak language equivalence, obtaining $A'$ and $B'$, then compose,
and minimise again. For minimisation, \texttt{Penrose} adapts algorithms recently proposed by Mayr and Clemente in~\cite{Mayr2013}, which does not guarantee optimality but is quite efficient in practice.
\item At each point, check for language emptiness; if at any point the language is empty then answer NO (not reachable). In the final step (corresponding to the root of the expression), non-emptiness is equivalent to reachability.
\end{enumerate}

In order to understand the algorithm, it is useful to run through a particular example execution. For the 3-bit counter, the input is the expression
\begin{equation}\label{eq:syntaxTree}
\cgr[height=7cm]{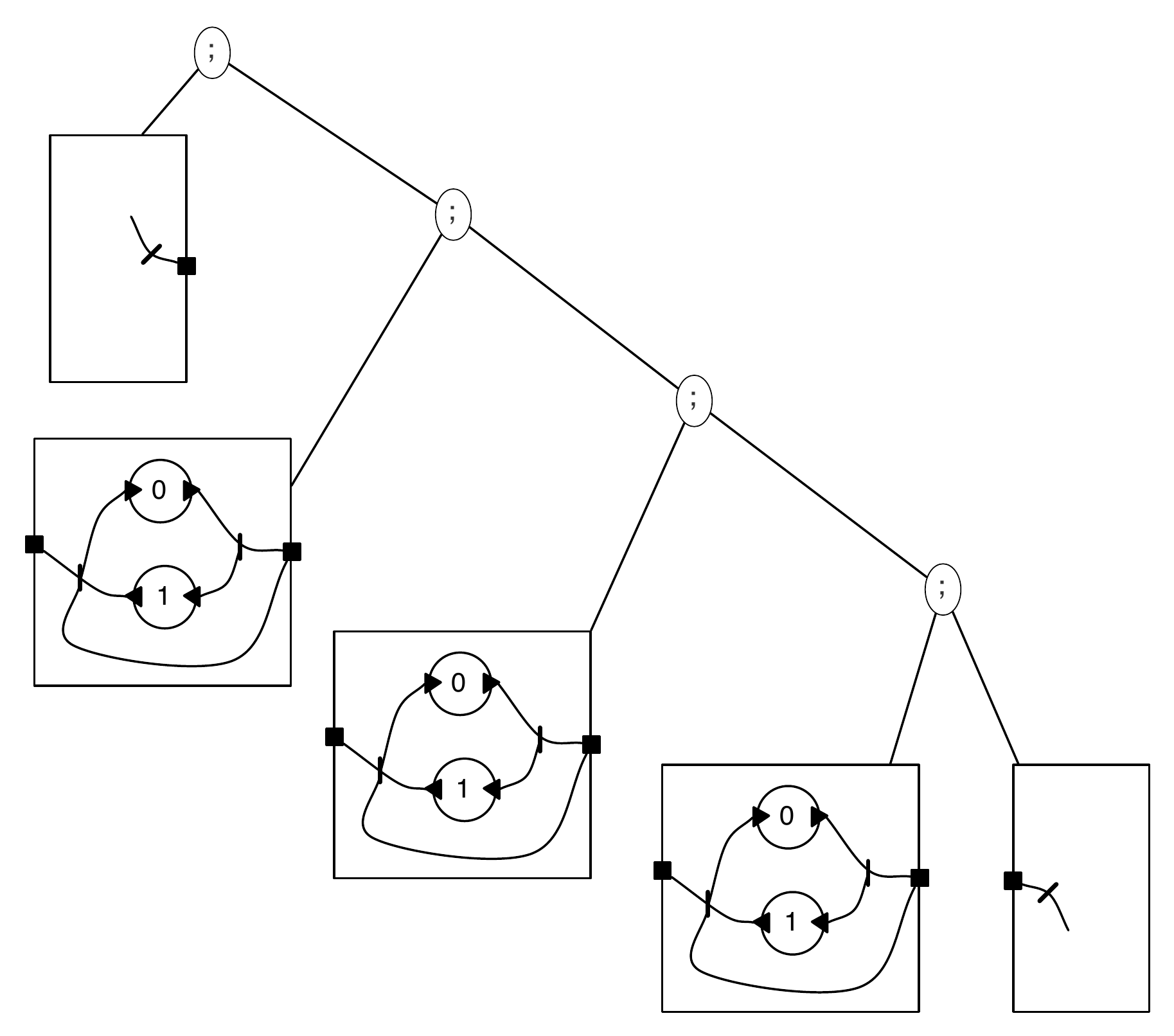}
\end{equation}
together with the information that we want to check the reachability of state $111$ from state $000$. Given this, we can transform the leaves to automata.
\[
\cgr[height=8cm]{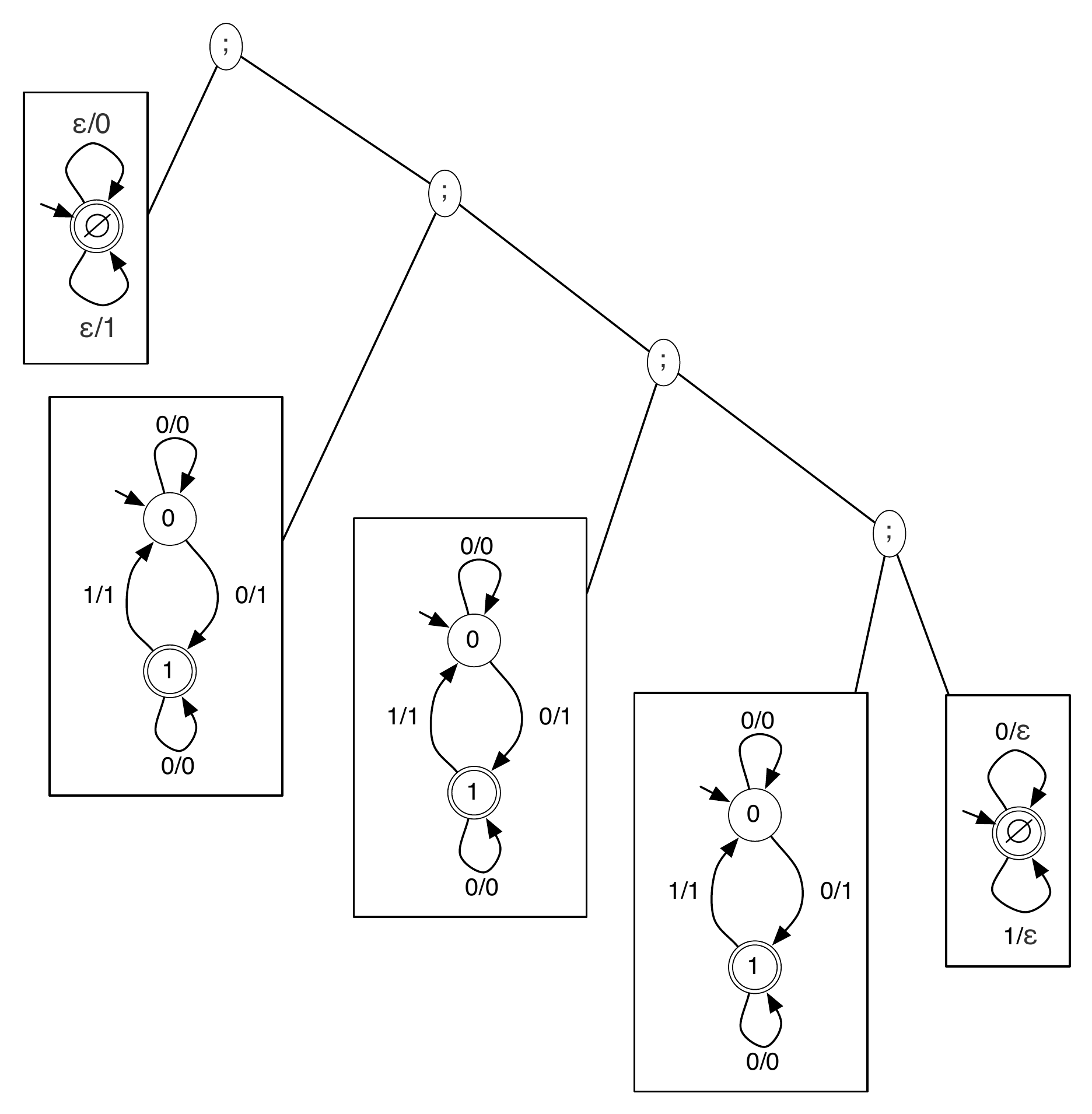}
\]
Performing the first, deepest composition and minimisation, as explained previously, yields 
\[
\cgr[height=6cm]{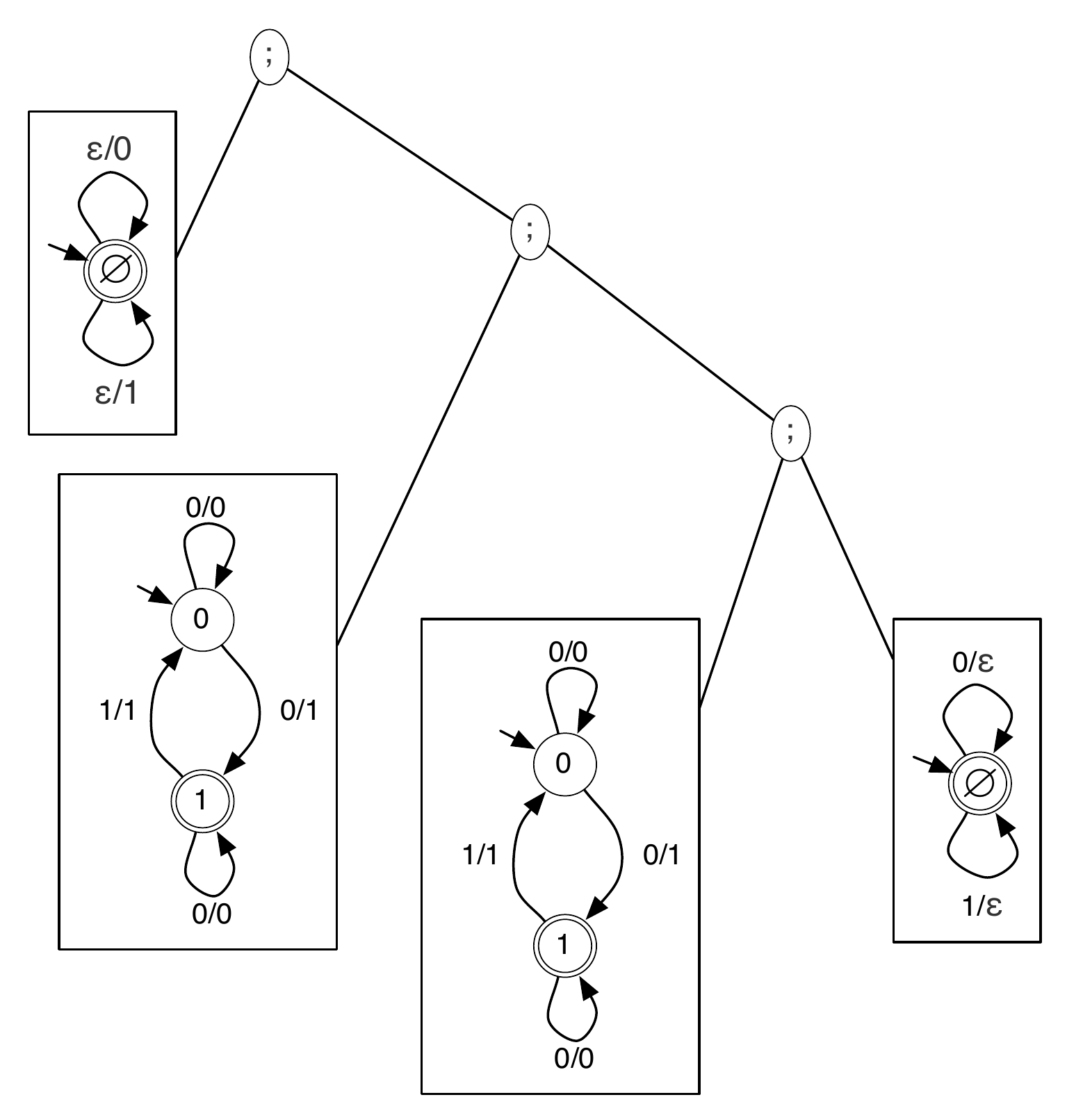}
\]
At this point \texttt{Penrose} uses memoisation to avoid doing the same (up-to language equivalence on the arguments) composition repeatedly. We use the recent advances in NFA language equivalence checking by Bonchi and Pous~\cite{Bonchi2013} to make this procedure efficient. 
At the final step we arrive at the trivially accepting automaton
\[
\cgr[height=2cm]{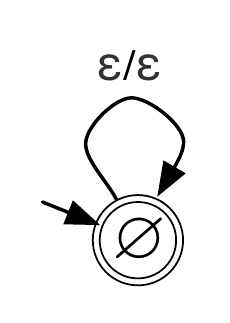}
\]
thereby verifying the reachability of $111$ from $000$. The combination of compositionality and weak language equivalence means that our problem, in which the execution path from the initial state to the final state is exponential in the size of the net, can be solved algorithmically in linear time.

The sceptical reader will no doubt wonder if this carefully chosen example is indicative of the utility of the technique in more realistic, practical examples. We have performed a compositional analysis on a standard set of benchmark examples. In several of these \texttt{Penrose}, unsurprisingly, beats its competitors which do not have access and do not take advantage of the high-level component-wise specification. See~\cite{Rathke2014a,Stephens2015} for a detailed account of the experimental results. In fact, many (most?) practical systems are quite modular by design, and thus consist of networks of repeated components: for many of such systems \texttt{Penrose} does very well, in several cases outperforming other tools by factors of magnitude.


\section{Future work}
\label{sec:future}



In recent years, the Bonchi, Zanasi and the author have studied the algebraic theory of signal flow graphs using props,
see e.g.~\cite{Bonchi2014b,Bonchi2015}. The algebraic theory of PNB can be studied using similar techniques, but the picture is both more interesting and more complex because one can study theories wrt a process equivalence such as language equivalence or bisimilarity. Even merely sound theories could be useful for model checking, since components could in principle be simplified as nets \emph{before} translating to automata.

Next, the algorithm presented in Section~\ref{sec:penrose} could be extended to deal with \emph{parametric} examples: e.g.\ instead of asking for the reachability in a 23-bit counter, one could ask the question in more generality for an $n$-bit counter, where the $n$ is a formal parameter. To achieve this, \texttt{Penrose} could be extended with logic that detects fixed points, such as the one encountered when evaluating the counter example---clearly, after the first composition and minimisation step, reachability has been decided for counters of arbitrary size. There are challenging related problems, including investigating heuristics and strategies for obtaining fixed points during evaluation.

A related, very challenging problem, and one that is currently completely side-stepped by \texttt{Penrose}, is obtaining efficient \emph{decompositions}, that is expressions such as~\eqref{eq:syntaxTree}, which are currently provided as input. Ideally, the process of finding the decomposition itself ought to be automated. Here there are connections with graph theory, and in particular the notion of \emph{rank-width}; some initial work has been done in~\cite{Rathke2013} and the connection with graph theory was made explicit in~\cite{Chantawibul2015}.

Finally, one could use the algebra of infinite state case PNBs to obtain approximations to reachability (and coverability) for infinite state nets; a decidable problem that it notoriously challenging for both theory and tool development.

\nocite{*}
\bibliographystyle{eptcs}
\bibliography{dcmdoi}
\end{document}

%% file: macro.tex
\newcommand{\Defeq}
 {\stackrel{\mathrm{def}}{=}}

\newcommand{\stran}{\raise1pt\hbox{$\centerdot$}}

\newcommand{\rring}[1]{\ensuremath{\mathbb{#1}}}

\newcommand{\N}{\rring{N}}

\newcommand{\ladj}[2]{\ar@/^/[#1]^-{#2} \ar@{}[#1]|-%

{\ifthenelse{\equal{#1}{r}}{\bot}{%

{\ifthenelse{\equal{#1}{rr}}{\bot}{%

{\ifthenelse{\equal{#1}{l}}{\top}{%

{\ifthenelse{\equal{#1}{u}}{\dashv}{%

{\vdash}}}}}}}}}}

\newcommand{\radj}[2]{\ar@/^/[#1]^-{#2}}

\newcommand{\radjff}[2]{\ar@{_{(}->}[#1]^{#2}}

\newcommand{\pullbacktop}[4]{%

{#1} \ar@/_/[ddr]_{#4} \ar@/^/[drr]^{#2}%

\ar@{.>}[dr]|-{#3} \\}

\newcounter{ncomm}

\newcommand{\ltsred}[1]
{ \setbox0=\hbox{$\ {}^{#1}\ $}
  \setbox1=\hbox{$\longrightarrow$}
  \loop\setbox1=\hbox{$-$\kern-0.3em\unhbox1}\ifdim\wd1<\wd0\repeat
  \hbox{$\ \ \mathop{\box1}\limits^{#1}\ \ $}
}

\newcommand{\arx}[2]{\!\xymatrix@=15pt{\ar[r]^{{#1}}_{{#2}}&}\!}

\newlength{\mylength}

{\setlength{\fboxsep}{15pt} 
\setlength{\mylength}{\linewidth}%
\addtolength{\mylength}{-2\fboxsep}%
\addtolength{\mylength}{-2\fboxrule}%
\Sbox 
\minipage{\mylength}%
\setlength{\abovedisplayskip}{0pt}%
\setlength{\belowdisplayskip}{0pt}%
}%
{\endminipage\endSbox 
\[\fbox{\TheSbox}\]}